\documentclass[conference]{IEEEtran}

\usepackage{cite}
\usepackage{amsmath,amssymb,amsfonts}
\usepackage{algorithmic}
\usepackage{graphicx}
\usepackage{textcomp}
\usepackage{booktabs}
\usepackage{xcolor}
\usepackage{subcaption}
\usepackage{placeins}
\usepackage{multirow}
\usepackage{bm}
\usepackage{hyperref}
\usepackage{academicons}
\usepackage{makecell} 
\usepackage{hyperref}
\usepackage{graphicx}

\makeatletter
\AtBeginDocument{\DeclareMathVersion{bold}
\SetSymbolFont{operators}{bold}{T1}{times}{b}{n}
\SetMathAlphabet{\mathrm}{bold}{T1}{times}{b}{n}
\SetMathAlphabet{\mathit}{bold}{T1}{times}{b}{it}
\SetMathAlphabet{\mathbf}{bold}{T1}{times}{b}{n}
\SetMathAlphabet{\mathtt}{bold}{OT1}{pcr}{b}{n}
\SetSymbolFont{symbols}{bold}{OMS}{cmsy}{b}{n}
\renewcommand\boldmath{\@nomath\boldmath\mathversion{bold}}}
\makeatother

\def\BibTeX{{\rm B\kern-.05em{\sc i\kern-.025em b}\kern-.08em
    T\kern-.1667em\lower.7ex\hbox{E}\kern-.125emX}}

\begin{document}

\title{Folded context condensation in Path Integral formalism for infinite context Transformers}

\author{
    \IEEEauthorblockN{
        Won-Gi Paeng\IEEEauthorrefmark{1}\href{https://orcid.org/0000-0001-8334-5343}{\includegraphics[scale=0.05]{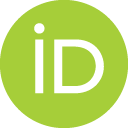}}, 
        Daesuk Kwon\IEEEauthorrefmark{1}\href{https://orcid.org/0009-0009-8621-5785}{\includegraphics[scale=0.05]{orcid.png}}, 
        Kyungwon Jeong\IEEEauthorrefmark{1}\href{https://orcid.org/0009-0009-0407-2953}{\includegraphics[scale=0.05]{orcid.png}}, 
        and Honggyo Suh\IEEEauthorrefmark{1}\href{https://orcid.org/0000-0003-0596-8873}{\includegraphics[scale=0.05]{orcid.png}}
    }
    \IEEEauthorblockA{\IEEEauthorrefmark{1}Hyntel, Inc., Seoul 04783, Korea}
    \IEEEauthorblockA{Corresponding author: Won-Gi Paeng, wgpaeng@hyntel.net}
}

\maketitle

\begin{abstract}
In this work, we present a generalized formulation of the Transformer algorithm by reinterpreting its core mechanisms within the framework of Path Integral formalism. In this perspective, the attention mechanism is recast as a process that integrates all possible transition paths leading to future token states, with temporal evolution governed by the Feed-Forward Network. By systematically mapping each component of the Transformer to its counterpart in the Path Integral formulation, we obtain a more compact and efficient representation, in which the contextual information of a sequence is condensed into memory-like segments. These segments are recurrently processed across Transformer layers, enabling more effective long-term information retention. We validate the effectiveness of this approach through the Passkey retrieval task and a summarization task, demonstrating that the proposed method preserves historical information while exhibiting memory usage that scales linearly with sequence length. This contrasts with the non-linear memory growth typically observed in standard attention mechanisms. We expect that this quantum-inspired generalization of the Transformer architecture will open new avenues for enhancing both the efficiency and expressiveness of future Transformer models.
\end{abstract}

\section{Introduction}

Transformers architecture~\cite{Vaswani} has revolutionized natural language processing (NLP) by demonstrating unparalleled effectiveness in capturing long-range dependencies and enabling scalable, parallelized computation. This architecture often consists of an embedding layer, decoder layers with a self-attention mechanism, and a Feed-Forward Network (FFN), therefore, it predicts the next token in an auto-regressive manner. Unlike earlier Recurrent Neural Network (RNN) based models, which process sequences sequentially and suffer from vanishing gradient issues~\cite{Bengio1994}, Transformers leverage self-attention mechanisms to establish global dependencies across tokens, leading to significant improvements in contextual reasoning and coherent text generation. This architectural advantage has propelled Transformers to the forefront of modern NLP, underpinning the development of large-scale language models (LLMs) with remarkable generative capabilities.

However, Transformers have suffered from their computational inefficiency when handling long sequences. The self-attention mechanism inherently scales quadratically with sequence length in both memory and computation, making it impractical to process extensive contexts within a single pass. This limitation hinders the retention of dependencies across long documents, dialogues, or continuous interactions, prompting the development of alternative strategies such as Infini-Attention~\cite{Munkhdalai}, which restructures the self-attention mechanism by introducing a hybrid memory system that allows interaction between local context tokens and external memory components, leveraging a linear attention~\cite{Shen2024} framework to reduce computational overhead. Another innovative solution is Titans~\cite{Ali}, which incorporates trainable neural long-term memory, enabling the model to retain and retrieve past contextual information beyond traditional Transformer constraints.

Despite their empirical success, moreover, the theoretical foundations of Transformer architecture remain incomplete. While extensive research has explored their scaling properties~\cite{Kaplan2020}, emergent behaviors~\cite{Wei2022}, and empirical performance~\cite{Srivastava2022, Hendrycks2020}, these studies provide limited insight into the fundamental mechanisms governing Transformer architectures. In particular, there is a lack of rigorous theoretical formulations explaining how information propagates through layers and how hierarchical structures are captured within Transformer models. This gap in understanding limits the interpretability and optimization of Transformer-based architectures and presents challenges in extending their capabilities systematically.

To address these challenges, we propose a novel theoretical generalization of the Transformer architecture based on the Path Integral formalism~\cite{Feynman1965}—a fundamental framework in quantum mechanics used to describe the evolution of dynamical systems. The Path Integral approach models a system's transition by summing over all possible trajectories, each weighted by an exponential factor of the action. Based on this principle, we reinterpret the attention mechanism in Transformers as integrating over all potential transition paths leading to future token states. This perspective allows us to establish a formal analogy between token state evolution in Transformers and the summation of quantum paths, with the transition dynamics governed by the Least-action principle.

Through this framework, we introduce a structured context delivery mechanism that enhances long-term dependency modeling while maintaining computational efficiency. By treating each Transformer layer as a time-evolution operator acting on token states, we derive a formulation in which contextual information is recurrently processed across layers. This approach effectively condenses past states into memory-like segments that evolve over time, thereby reducing memory complexity to scale linearly with sequence length while preserving expressive power. Unlike traditional self-attention, which explicitly computes pairwise token interactions within a sequence at each step, our method naturally integrates historical context in the separated segments of a sequence without incurring quadratic computational costs.

Building on this theoretical foundation in Section \ref{gPath-inte}, we present an implementation in Section \ref{implementation1} and \ref{implementation2} that extends Transformer architectures with a time-evolving memory buffer. This structured approach ensures that past token representations contribute constructively to future states, akin to propagators in quantum field theory. Our proposed model enables efficient long-range dependency modeling while significantly reducing memory consumption compared with conventional Transformers.

In section \ref{exp}, we validate our approach through empirical evaluations, including the Passkey retrieval and summarization tasks. Our experimental results demonstrate that the proposed method achieves performance on par with the base model in contextual recall while delivering superior memory efficiency compared with the base Llama-3.2 models~\cite{Aaron}.

By unifying Transformer architectures with principles from quantum mechanics, this work provides a novel theoretical perspective that enhances both interpretability and efficiency. We anticipate that this Path Integral-based formulation will open new avenues for developing scalable and theoretically grounded AI models, enabling more robust processing of long-range dependencies in natural language and beyond.

\section{Generalization of Transformer in Path Integral framework} \label{gPath-inte}
\begin{figure*}[t!]
    \centering  
    \includegraphics[width=0.9\linewidth]{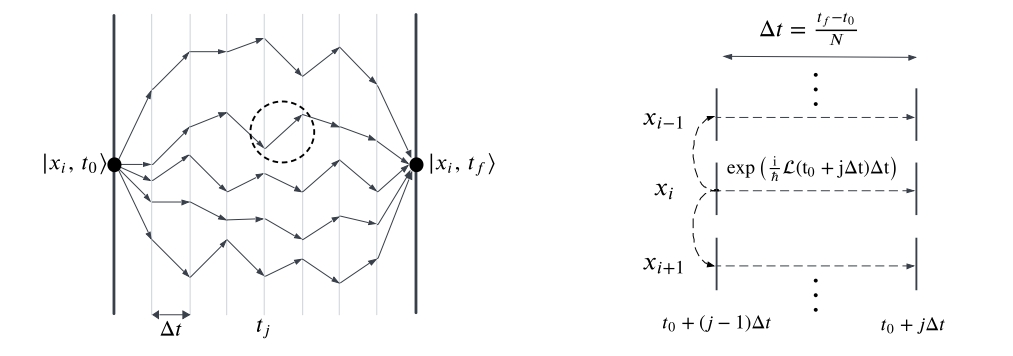}  
    \caption{Visualization of token state evolution in the Path Integral framework.
    The state \( |x_i, t_0\rangle \) at time \( t_0 \) evolves through multiple interaction pathways before reaching \( |x_i, t_f\rangle \)(left side). 
    Each evolution from $t_j$ to $t_j + \Delta t$ can be described as the transitions to all possible token states and the temporal evolutions of the projection states(right side).}
    \label{fig1}
\end{figure*}

The Path Integral approach provides a probabilistic framework for describing the evolution of a particle’s state by summing over all possible trajectories, with each path contributing according to a phase factor determined by the action. In classical mechanics —where the Planck constant $h$(=$2\pi\hbar$) is negligible with respect to the system’s characteristic energy scale— phase factors of the form $\exp(iS/\hbar)$ oscillate rapidly over time and cancel out for all but those paths near the classical trajectory. Consequently, only the trajectory that minimizes the action survives, so that the system’s motion follows the Principle of Least Action. Thus, the Path Integral formulation extends the Least Action Principle by incorporating every conceivable path to a future state, whose probability amplitudes interfere constructively or destructively. Here, the ratio of the system’s energy (or action) scale to the Planck constant determines whether its evolution appears essentially deterministic or fundamentally probabilistic.

Building upon this interpretation, we propose a novel generalization of the Transformer mechanism using the Path Integral formalism. In standard Transformers, the attention operation first determines the relationships between tokens within a sequence, followed by the FFN processing this contextualized information. This process bears a strong resemblance to the way quantum states evolve over time through the summation over all possible paths. By drawing on this analogy in the left side of Fig. \ref{fig1}, we reinterpret the transformation of a token vector through Transformer layers as the time evolution of a token state, where each token traverses multiple interaction pathways before contributing to the final output. Specifically, when we consider a case that a token state $|x_i,\, t_0 \rangle$ at position \( x_i \) in a sequence and time \( t_0 \) temporally evolves to the state $|x_i,\, t_{\rm f} \rangle$ at position \( x_i \) and time \( t_{\rm f} \), we can formulate this transition within a Path Integral framework as follows: 
\begin{align}
|x_i,\, t_{\rm f}\rangle = {\cal N} \sum_{\rm{all}\,\, \rm{paths}} {\cal T} \exp\left( \frac{i}{\hbar}\int^{t_{\rm f}}_{t_{0}} {\cal L} \,dt  \right) |x_i,\, t_{0}\rangle\, \label{att1}
\end{align} where ${\cal L}$ is the Lagrangian governing the dynamics (analogous to the action functional in classical mechanics), ${\cal N}$ is overall normalization factor to the contribution of all possible paths and ${\cal T}$ is the operator of the time-ordered product. Here, it should be noted that $\hbar$ does not correspond to the physical value of the Planck constant, but rather functions as a tunable parameter reflecting the characteristic energy scale of the token system under consideration.
In the condition $ \Delta t = \frac{t_{\rm f}-t_0}{N}\rightarrow 0$, (\ref{att1}) can be rewritten as
\begin{align}
|x_i,\, t_{\rm f}\rangle &= {\cal N} \sum_{\text{all paths}} {\cal T} 
\Bigg\{ \prod_{j=1}^{N} 
\exp\left( \frac{i}{\hbar} {\cal L}\left( t_0 + j\Delta t \right) \Delta t  \right) 
\Bigg\}  \notag \\  
&\quad | x_i,\, t_0\rangle\,.
\label{att2}
\end{align}

To consider the all possible paths in the way from $t_0$ to $t_{\rm f}$, we input the identity matrix as
\begin{equation}
I_j = {\cal N}_{j}\sum_{k} |x_k,\, t_{j}\rangle \langle x_k,\, t_{j} |
\end{equation} with the normalization factor ${\cal N}_j$ at each time $t_{j} \equiv t_0 + \left(j-1 \right) \Delta t$, which supposes transitions to the all possible token states at $t_{j}$. Here, we note that $x_k$ means $k$-th token position which is different from the ordinary position in space-time.
Then, finally we get
\begin{align}
|x_i,\, t_{\rm f}\rangle &= {\cal N} {\cal T} \bigg\{ \prod_{j=1}^{N} \Bigg[\exp\left( \frac{i}{\hbar} {\cal L}\left( t_0+j\Delta t \right) \Delta t  \right) \notag \\
&\quad {\cal N}_j \sum_{k} |x_k,\, t_{j} \rangle \langle x_k,\, t_{j}| \Bigg] \bigg\} | x_i,\, t_0\rangle\,. 
\label{att2}
\end{align}

In particular, we take into account the time evolution of the state from $t_0+(j-1)\Delta t$ to $t_0+j\Delta t$, from (\ref{att2}) as
\begin{align}
|x_i,\, t_{j+1} \rangle = \exp\left( \frac{i}{\hbar} {\cal L}\left( t_j+\Delta t \right) \Delta t  \right) \notag \\
\quad {\cal N}_j \sum_{k} |x_k,\, t_{j}\rangle \langle x_k,\, t_{j} |x_i,\, t_{j}\rangle \label{att3}
\end{align} which is illustrated in the right side of Fig. \ref{fig1}.

From (\ref{att3}), we induce the attention and neural net structure of the $j$-th Transformer layer as
\begin{align}
X_{i}^\prime &= {\rm{softmax}} \left[ \frac{X_{i} W_Q W_K^T}{\sqrt{d_M}}
\begin{bmatrix}
\cdots & X^{T}_{k-1} & X^{T}_{k} & \cdots
\end{bmatrix}
\right] \notag \\
&\qquad \left( \begin{matrix}
\cdots \\ X_{k-1} \\ X_{k} \\ \cdots
\end{matrix} \right)
W_V W_{FF} \,. 
\end{align} for the position and token embedded state on the position $i$, $X_i \in \mathbb{R}^{\,d_{E}}$, transforming to the context vector $X_i^\prime$ of the Transformer
by the following correspondence,
\begin{align}
| x_i,\, t_j \rangle
&\Leftrightarrow& Q^T = W_Q^T X_i^T \\
\langle x_k,\, t_j|
&\Leftrightarrow& K = X_k W_K \\
| x_k,\, t_j \rangle
&\Leftrightarrow& V^T = W_V^T X_k^T  \\
\exp\left( \frac{i}{\hbar} {\cal L}\left( t_j+\Delta t \right) \Delta t  \right)
&\Leftrightarrow& W_{FF}^T \\
|x_i,\, t_{j+1} \rangle
&\Leftrightarrow& X_i^{\prime\, T}\,,
\end{align} where $FF$ stands for Feed-Forward Network and $W_{Q,\,K,\,V} \in \mathbb{R}^{\,d_E\times d_M}$ are query, key and value projection matrices. That is the generalized form of decoder-only Transformer. Here we ignored multi-head attention for simplicity.

Training the Transformer can then be understood as an optimization process that identifies the paths near the Least Action within this Path Integral formulation. Specifically, minimizing the cross-entropy loss between the predicted outputs and target tokens can be interpreted as selecting the optimal trajectories that satisfy the Principle of Least Action. Mathematically, this is expressed as
\begin{equation} -\ln \left| \mathcal{M}_{ij} \right|^{2} = -\ln \left| \langle i | {\cal N} \sum_{\rm{all}\,\, \rm{paths}} {\cal T} \exp\left( \frac{i}{\hbar}\int {\cal L} \,dt \right) | j \rangle \right|^2\,. \end{equation}
Here, the transition amplitude $\mathcal{M}_{ij}$ represents the probability amplitude of moving from the initial state $|j\rangle$ to the final state $|i\rangle$, summing over all possible evolutionary pathways governed by the Lagrangian $\mathcal{L}$. The Transformer training process thus naturally suppresses high-frequency paths that contribute minimally to the model's final output while reinforcing trajectories that minimize the action. This mechanism leads to the emergence of the Least Action Principle in Transformer optimization:
\begin{equation}
-\ln \left| \mathcal{M}_{ij} \right|^{2} \rightarrow 0\,.
\end{equation}

Building on the above correspondence, we refine our interpretation of Transformers by viewing each Transformer layer as a temporal slice through which all possible computational paths evolve toward the final output state. Each layer, composed of an FFN and an attention mechanism, represents a discrete point in this evolution. A token state undergoes temporal evolution each time it passes through the FFN, resulting in a phase shift of the state. The attention mechanism, in contrast, evaluates the contextual relevance among all input token states. If the phase difference between two tokens leads to destructive interference, these token states are less likely to exchange contextual information. Therefore, the temporal evolution of a token —particularly the associated phase shift induced by the FFN— needs to be carefully treated in inter-token interactions. It is important to note that tokens located at different positions within a sequence do not necessarily correspond to distinct temporal moments. In this formulation, time is not defined by positional indices but emerges through the accumulated phase shifts governed by the FFN at each layer, leading to the final output. Accordingly, the primary factor to consider is the relative phase difference between token states situated at different positions. 

In the following section, we explore the practical implications of this framework for modeling how contextual information can be preserved across segments that are input into Transformer layers in sequential orders.

\section{Folded context Condensation}\label{implementation1}

\begin{figure}[h]
    \centering
    \includegraphics[width=0.45\textwidth, height=7cm]{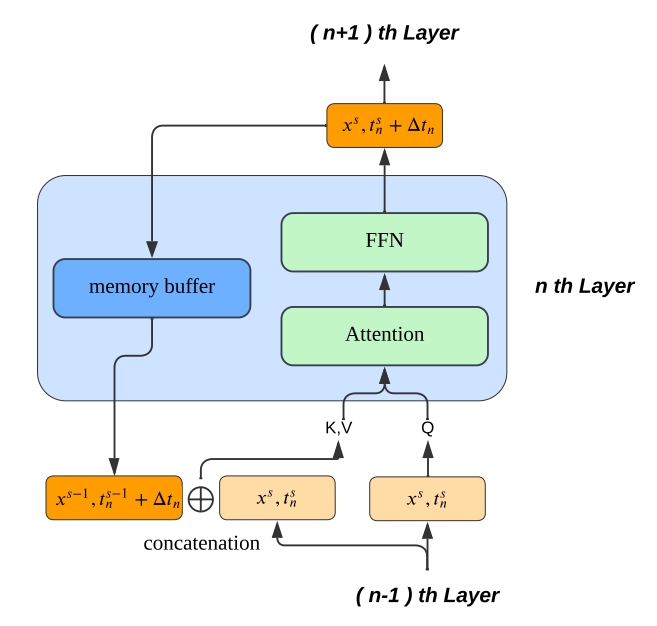}
    \caption{Underlying architecture of the model. A segment $|x^s,\, t_n^s \rangle$ from the previous layer is projected into a query and also concatenated with time-evolved memory buffer $|x^{s-1},\, t_n^{s-1}+\Delta t_n \rangle$ to be key and value. The output of the current layer will be saved as a memory buffer and retrieved when the next segment arrives.}
    \label{fig:structure}
\end{figure}

In our Transformer architectures, token states undergo temporal evolution governed by the FFN across layers, leading to phase shifts that influence coherence in long-range dependencies. If unregulated, these shifts may induce destructive interference between token states of separate segments, causing information loss. To address this, we analyze the impact of token state phase shifts and introduce a periodic phase gap mechanism that ensures constructive interference. 

By rewriting (\ref{att3}), the token state at position $i$ in segment $s$ evolves temporally through the FFN at each Transformer layer as
\begin{align}
|x_i^s, t_{n+1}^s \rangle &=  \exp\left( \frac{i}{\hbar} {\cal L} \left( t_{n}^s +\Delta t \right) \Delta t  \right) \notag \\
&\quad {\cal N} \sum_{k} |x_k^s, t_n^s \rangle \langle x_k^s, t_n^s | x_i^s, t_{n}^s\rangle\,, \label{evolution}
\end{align} where we set the time of input segment $s$ into $n$-th layer as $t_n^s$. Here, we denote that the time of a token state represents the phase of the state, so the states sharing same time exhibit constructive or zero phase difference between them.
This equation suggests that each Transformer layer, after training, exhibits distinct phase evolution profiles aligning with eigenmodes of token state transitions. Each layer parameters adapt to minimize phase misalignment, leading to the emergence of layer-specific eigen-frequencies.

This phenomenon is analogous to resonant systems such as stringed or wind instruments, where standing wave patterns form due to imposed boundary conditions. Consequently, the phase evolution of token states given in (\ref{evolution}) follows the wave equation:
\begin{equation}
i\hbar\frac{\partial}{\partial t} \Psi_i^s - \mathcal{H} \Psi_i^s = 0,
\end{equation}
where $\Psi_i^s$ denotes the token state at position $i$ in segment $s$, and $\mathcal{H}$ represents the Hamiltonian of the layer’s transformation governed by the Lagrangian. This equation leads to standing wave solutions constrained by the boundary conditions between input and output tokens, indicating that token states evolve according to intrinsic phase shifts governed by learned parameters. These parameters are trained to align phase shifts with resonant frequencies, $f_k$'s, at each layer:
\begin{equation}
\Delta \phi_n = \omega_n(\cdots,\, f_k, \cdots) \Delta t_n, \quad \omega_n \sim F(\mathcal{H},\,\Psi_i^s),
\end{equation}
where $\omega_n$ is the overall frequency, a function of resonant frequencies, and given by the input token state $\Psi_i^s$ applied by the Hamiltonian. We note that $\Delta t_n$ is the time period of the standing wave at the $n$-th layer, and each layer has its own eigen time period. After training, each time period at each layer is fixed to ensure that token state transitions reinforce the context established by prior layers rather than interfere destructively.

To prevent destructive interference, we introduce a periodic phase gap between segments, enforcing coherence in token state transitions. Misaligned phase differences between adjacent segments can introduce interference effects, hampering information retention. We define the segment-to-segment transition getting from (\ref{evolution}) as
\begin{align}
|x_i^{s}, t_{n}^s +\Delta t_n\rangle &= \exp\left( \frac{i}{\hbar} {\cal L} \left( t_{n}^s +\Delta t \right) \Delta t  \right) \notag \\
&\quad {\cal N} \sum_{k} \left( \left|x_k^{s-1}, t_{n}^{s-1} + \Delta t_n \rangle \langle x_k^{s-1}, t_{n}^{s-1} + \Delta t_n \right| \right. \notag \\
&\qquad \left. + \left|x_k^{s}, t_{n}^{s} \rangle \langle x_k^{s}, t_{n}^{s} \right| \right) | x_i^{s}, t_{n}^{s}\rangle.
\label{segmenttosegment}
\end{align}
By explicitly incorporating prior segment states into the transition, the alignment of segments at integer-multiple time gaps,
\begin{equation}
t_n^s = t_{n}^{s-1} + \Delta t_n,
\end{equation}
enables constructive interference, preserving historical context across segments.

To implement this in practice, we propose a recurrent segment processing approach wherein token states from previous segments are periodically reinjected into the Transformer as shown in Fig. \ref{fig:structure}. This structured recurrence ensures phase-aligned reinforcement, analogous to standing wave resonances. By enforcing these phase-aligned boundary conditions, the architecture sustains long-range dependencies through constructive interference, mitigating contextual degradation over extended sequences.

\section{The implementation of the model}\label{implementation2}

\begin{figure}[h]
    \centering
    \includegraphics[width=0.3\textwidth, height=3cm]{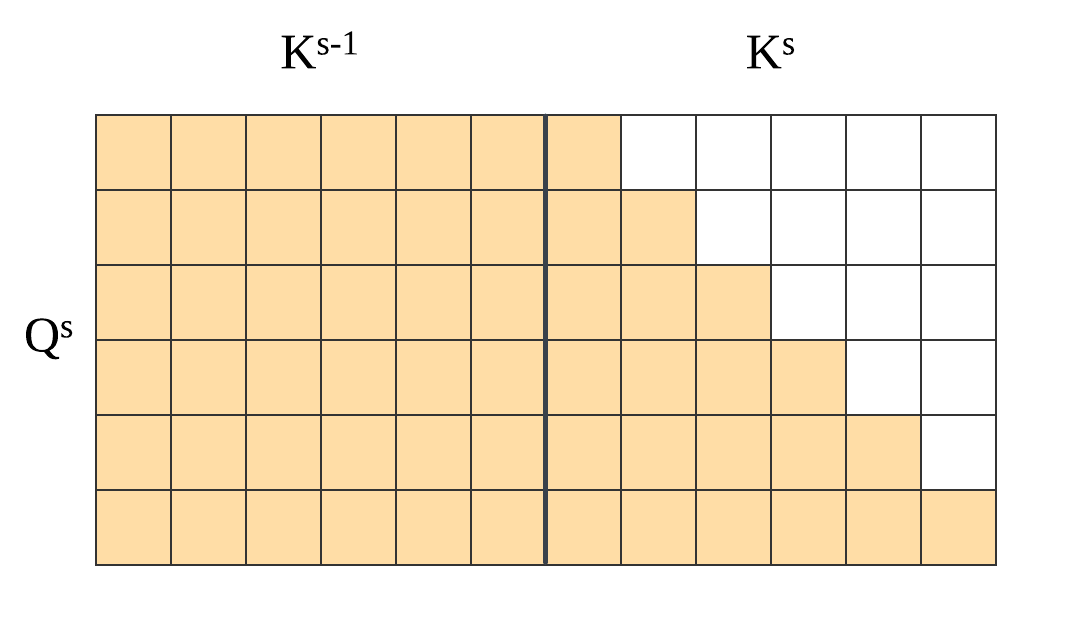}
        \caption{The attention mask is divided into two regions: $K^{s-1}$ represents memory keys that allow full attention across all positions, while $K^{s}$ represents current keys, where attention is restricted with a causal masking pattern. $Q^s$ denotes the current query, interacting selectively with both $K^{s-1}$ and $K^{s}$.}
    \label{fig:mask}
\end{figure}

Building upon the structured recurrence and phase coherence mechanisms introduced in the previous sections, we now formalize the implementation of segmental processing in Transformer architectures. Our approach optimizes memory efficiency while preserving long-range dependencies by leveraging recurrent attention across segment transitions.

To ensure seamless information propagation across segments, we define the periodic temporal evolution of token states in the Transformer layer as:
\begin{equation}
X^{s\, \prime} = \rm{softmax} \left( \frac{Q^s}{\sqrt{d_M}}
\begin{bmatrix}
K^{s-1\,T} & K^{s\,T}
\end{bmatrix}
\right) \left( \begin{matrix}
V^{s-1} \\ V^{s}
\end{matrix} \right)
W_{FF},
\end{equation}
where $Y^s = X^{s} W_Y$ with $X^s \in \mathbb{R}^{l\times d_{E}}$ and $Y = Q,\, K,\, V$. The attention mask structure, depicted in Fig. \ref{fig:mask}, ensures that the query $Q^s$ retains full access to past segment keys $K^{s-1}$ while maintaining causal restrictions within the current segment $K^s$. By enabling unmasked attention between $Q^s$ and $K^{s-1}$, we ensure that past contextual information remains accessible, reinforcing coherence between segments.

\begin{figure}[h]
\begin{center}
\includegraphics[width=0.5\textwidth, height=5.0cm]{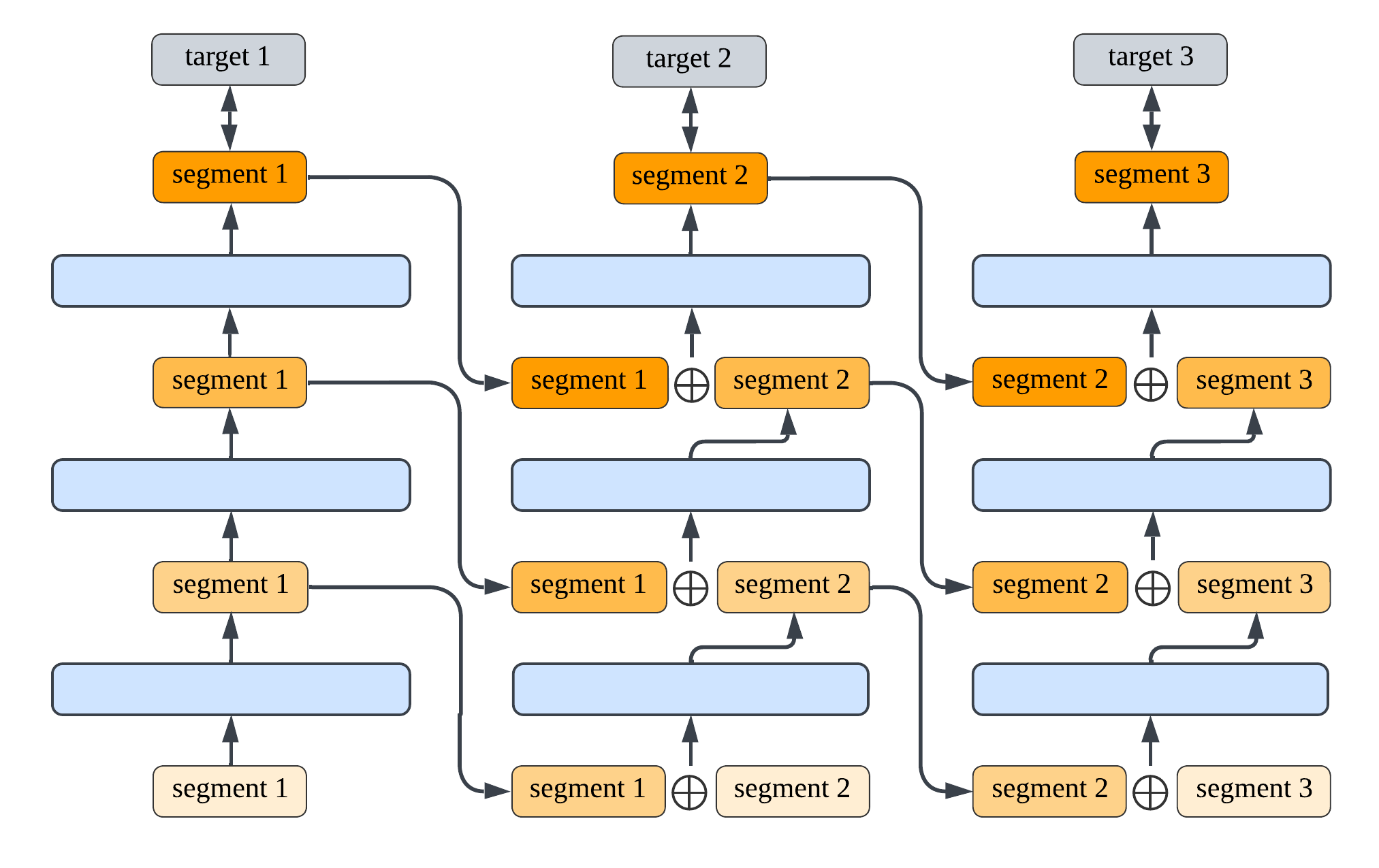}

\caption{Each output of each layer for the input segment is preserved as a memory buffer. Then, the memory buffer is concatenated with the current segment and takes part in the attention.}
\FloatBarrier

\label{evol1}
\end{center}
\end{figure}

Given that self-attention mechanisms scale quadratically with input length, reducing computational overhead is crucial for efficient training and inference. Our recurrent segment-wise approach mitigates the resource constraints associated with large-scale Transformers while ensuring robust sequence modeling. We achieve this by partitioning an input sequence of length $L$ into $N_s$(= $L/l$) segments, each of uniform length $l$, and sequentially processing them through the multi-layer Transformer as shown in Fig. \ref{evol1}.

During training, token embeddings are first assigned distinct positional indices per segment. The $s$-th segment is then concatenated with the recurrent representation from the $(s-1)$-th segment, forming a combined key-value set of length $2l$, while the query remains restricted to $l$ positions, as described in Fig. \ref{fig:structure}. This formulation effectively captures and propagates historical information while limiting redundant computation.

Following self-attention, the output undergoes residual connection processing and is passed through an FFN. The final layer-wise outputs align with corresponding segment-wise targets, yielding the total loss function:
\begin{equation}
L_{\rm{total}} = \sum_{s=1}^{N_s} L_s,
\end{equation}
where $L_s$ represents the cross-entropy loss for the $s$-th segment. The model is optimized to minimize $L_{\rm{total}}$, ensuring robust long-range dependency retention while maintaining computational efficiency.

\section{Experiments} \label{exp}
In this section, we evaluate the effectiveness and efficiency of the proposed Folded Context Condensation mechanism (hereafter referred to as the \textbf{Condensation model}) in comparison to the Llama-3.2 model (hereafter referred to as the Llama model). First, we demonstrate that the Condensation model achieves significantly lower memory consumption than the Llama model under comparable conditions, and we further evaluate its training speed across varying sequence lengths to assess whether the Condensation mechanism offers practical computational advantages. Next, we validate the Condensation model’s ability to propagate information across distant segments using a passkey retrieval task, which tests whether critical information from earlier segments can be effectively retrieved after processing subsequent segments. Finally, we evaluate the ability to perform long-document summarization using the Booksum dataset, showing that the model is capable of processing extended contexts and generating coherent summaries without quality degradation, even as the input length increases. At the end of this section, we will demonstrate through attention score visualization that, with Folded Context Condensation, the contextual information from preceding segments is not lost and continues to participate in the attention process together with the currently processed segment.

\subsection{Memory usage measurement \& Speed Comparison}
Compared to the original Transformer model, the Condensation model splits a sequence of length \(L\) into \(N_s\) fixed-length segments. This segmentation reduces the memory complexity of the attention computation from \(\mathcal{O}(L^2)\) to \(\mathcal{O}(N_s l^2)\), where \(l\) denotes the segment length. By fixing the segment length, the overall memory requirement transitions from quadratic to linear growth with respect to the total sequence length. The effectiveness of this approach was validated through direct comparison with the Llama model.

\begin{figure}[h]
    \centering
    \includegraphics[width=0.45\textwidth, height=5cm]{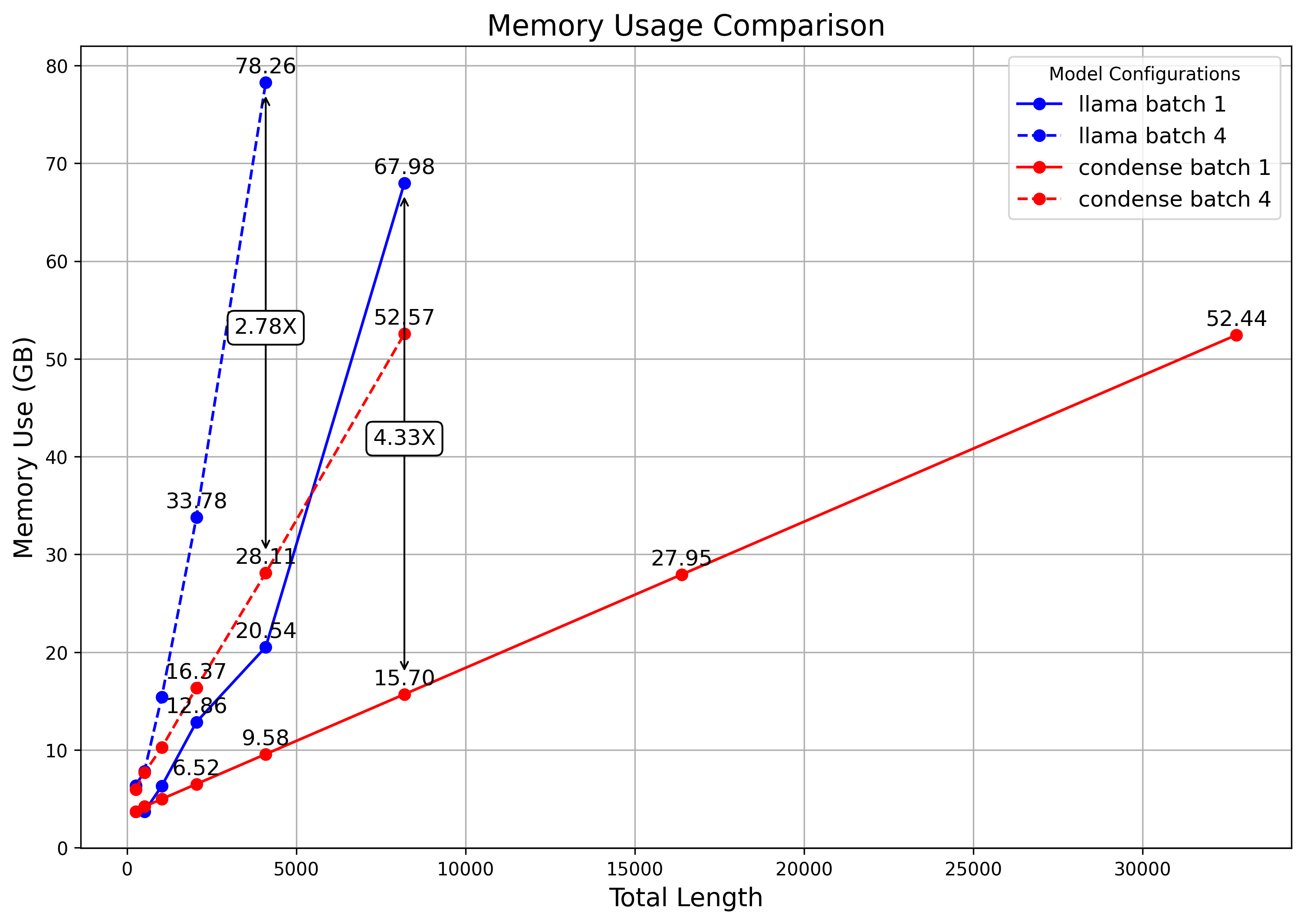}
    \caption{Memory usage comparison as a function of total sequence length. The plot illustrates memory consumption (in GB) for Condensation and Llama models across batch sizes 1 and 4.}
    \label{fig:memory_no_flash}
\end{figure}
\FloatBarrier
The Condensation model, represented by red lines in Fig.~\ref{fig:memory_no_flash}, exhibits a linear increase in memory usage proportional to the sequence length. In contrast, the Llama model, shown with blue lines, deviates from this linear trend, indicating a more complex memory scaling behavior. At a total sequence length of 8 K (K represents 1024), the Llama model consumed 4.33 times more memory compared to our Condensation model with batch size of 1. Similarly, at 4 K, the Llama model required 2.78 times more memory than the Condensation model with batch size of 4.\\

\begin{figure}[h]
    \centering
    \includegraphics[width=0.45\textwidth, height=5cm]{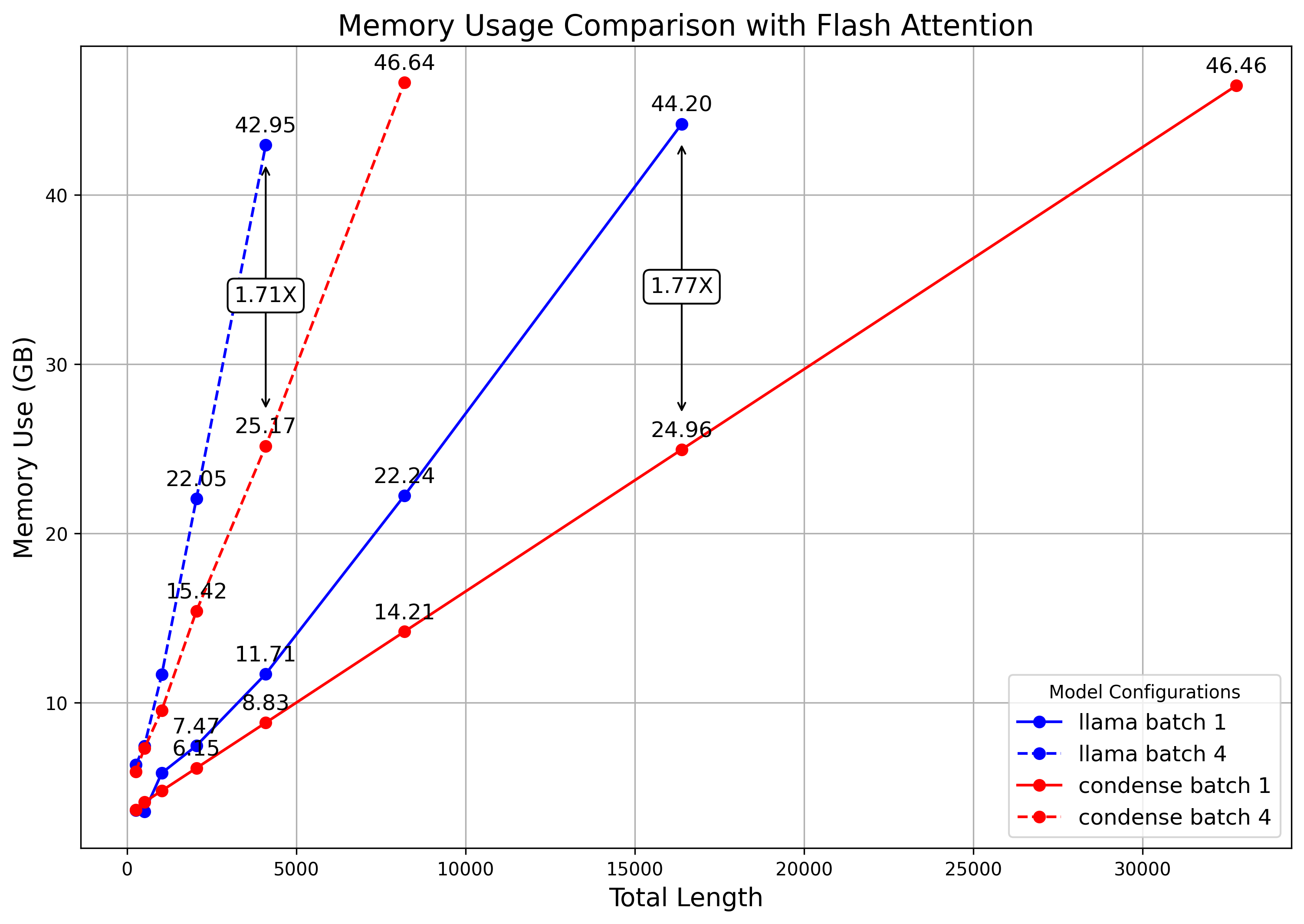}
    \caption{Memory usage comparison by sequence length with FlashAttention.}
    \label{fig:memory_flash}
\end{figure}
\FloatBarrier

Fig. \ref{fig:memory_flash} presents a comparison of memory usage as a function of total sequence length with FlashAttention \cite{Dao} enabled. At a sequence length of 16 K, the Llama model consumed approximately 1.77 times more memory than the Condensation model with a batch size of 1. Similarly, at a sequence length of 4 K, the Llama model required approximately 1.71 times more memory than the Condensation model when using a batch size of 4.

\begin{figure}[h]
    \centering
    \includegraphics[width=0.45\textwidth, height=5cm]{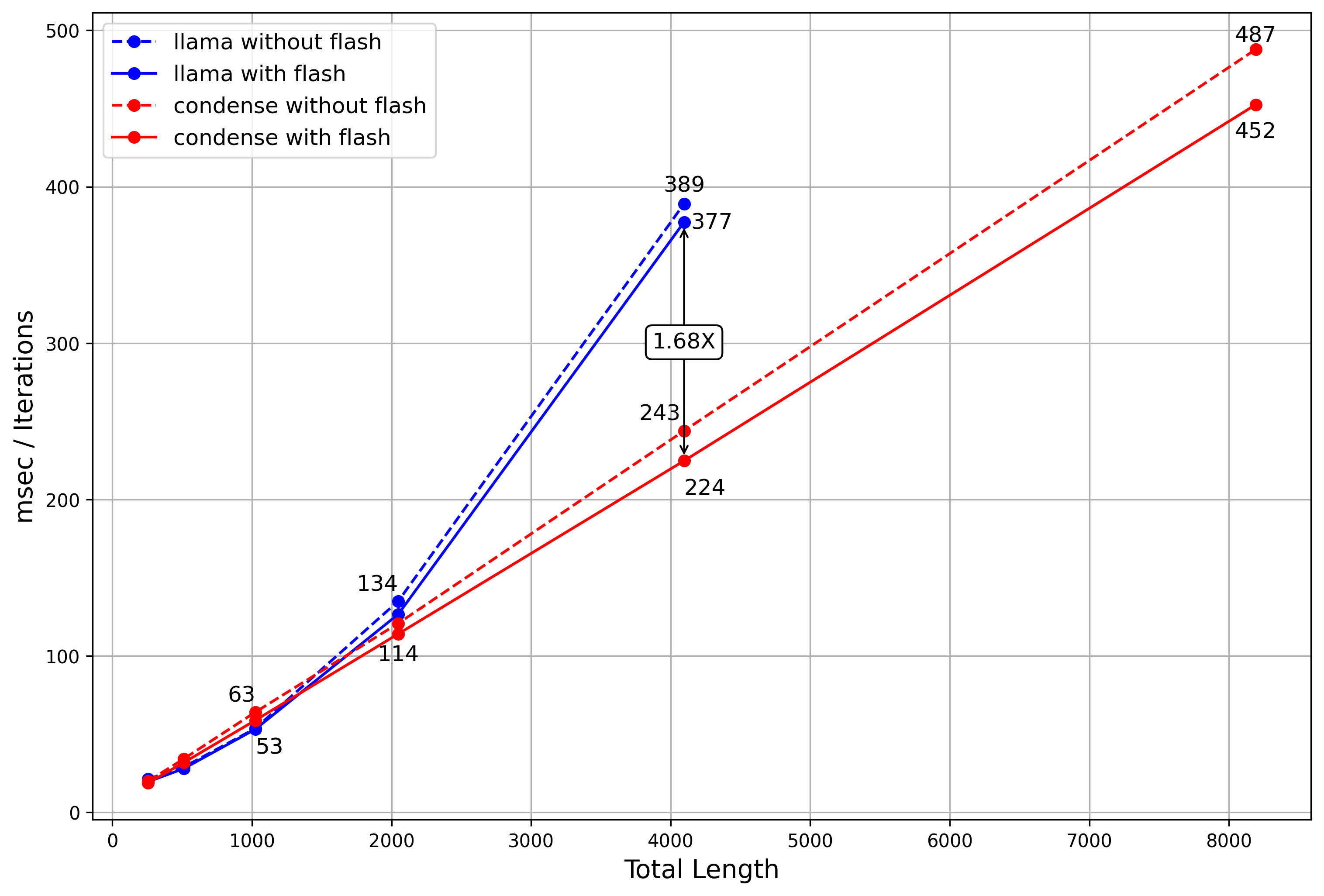}
    \caption{Training Speed Comparison by Sequence Length with and without FlashAttention}
    \label{fig:speedtotal.png}
\end{figure}

\FloatBarrier
Fig. \ref{fig:speedtotal.png} presents a comparison of training speeds between the Condensation and Llama models across varying sequence lengths, evaluated both with and without FlashAttention. For shorter sequences (under 2 K tokens), the Llama model achieves faster training speeds than the Condensation model, regardless of whether FlashAttention is applied. However, as sequence length increases beyond 2 K tokens, the Condensation model outperforms Llama in terms of speed, particularly for sequences of 4 K and 8 K tokens, where Condensation achieves lower iteration times. At a sequence length of 4 K, Condensation with FlashAttention processes training iterations approximately 1.68 times faster than Llama with FlashAttention. Also, the results demonstrate that the effect of FlashAttention is relatively minor for both models across all sequence lengths, with only marginal improvements observed in training speed.

\begin{table}[h!]
\centering
\begin{tabular}{@{}lcc@{}}
\toprule
\textbf{Configuration}       & \textbf{Condensation}& \textbf{Llama} \\ \midrule
Original model                        & Llama-3.2-1B         & Llama-3.2-1B   \\
Number of Layers             & 4                    & 4              \\
Hidden size                  & 512                  & 512            \\
Segment Length               & 256                  & -              \\
Trainable Parameters         & 126 M                & 126 M          \\
Head Dimension               & 64                   & 64             \\
Number of Heads              & 32                   & 32             \\
Number of Key/Value Heads    & 8                    & 8              \\ \midrule
Number of GPUs               & 1                    & 1              \\
GPU                          & H100                 & H100           \\ \midrule
\end{tabular}
\caption{Configuration for memory \& speed experiment}
\label{table:memoryandspeedconfig}
\end{table}

The memory and speed experiments were conducted using the configurations shown in Table \ref{table:memoryandspeedconfig}, with both the Condensation and Llama models tested on a single NVIDIA H100 GPU. In our experiments, we selected a segment size of 256 and 512 tokens for the Condensation model. This range was optimal given our resource constraints, requiring minimal additional GPU memory. While larger segment sizes could improve performance, we prioritized evaluating the model's ability to transfer information across segments, necessitating multiple segment blocks. To compare the impact of the Condensation mechanism on memory usage and training speed, we used a reduced-scale model with 126 million trainable parameters.

The results indicate that, in terms of memory efficiency, the Condensation architecture effectively reduces memory usage by flattening the memory growth curve, enabling the processing of significantly longer sequences with reduced memory overhead. In terms of training speed, however, the Condensation model exhibits slower iteration times for shorter sequences compared to Llama. As the sequence length increases beyond 2 K tokens, the Condensation model becomes more efficient in terms of speed, demonstrating faster iteration times than Llama at 4 K and 8 K tokens. In the current implementation, the Condensation model processes input sequences sequentially at the granularity of each segment, which limits the potential for significant speed improvements. However, future enhancements incorporating pipeline-style parallelism could further improve Condensation's processing efficiency, particularly for long sequences.

\subsection{Passkey retrieval task}
\begin{figure}[h]
    \centering
    \hspace*{-0.5cm} 
\includegraphics[width=0.5\textwidth, height=8cm]{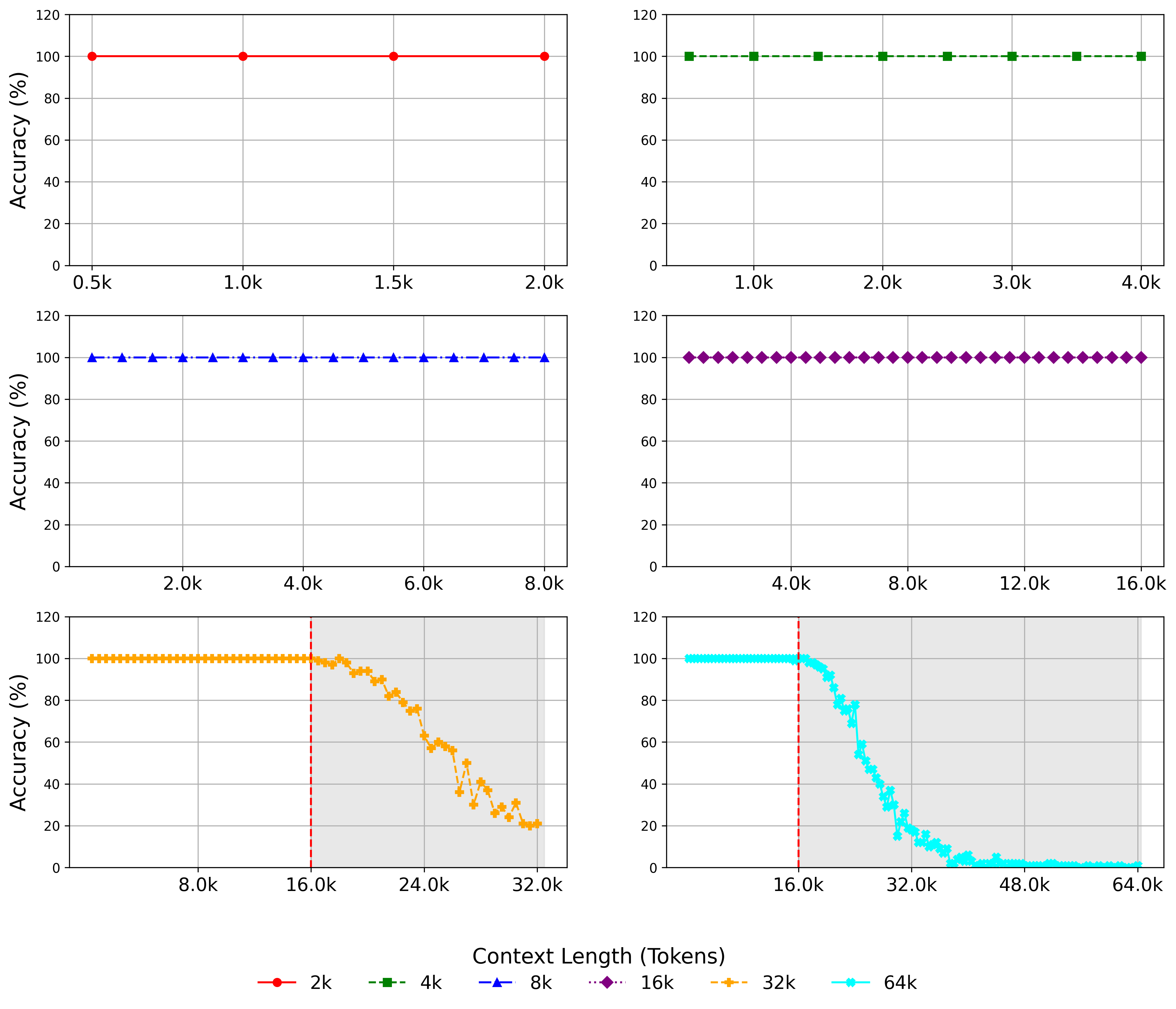}
    \caption{Passkey retrieval accuracy across different total sequence lengths. Each subplot represents evaluation at a specific context length, ranging from 2 K to 64 K tokens. The vertical dashed red line indicates the sequence length at which the model was fine-tuned. Accuracy remains consistently high within the fine-tuned length, while performance degrades when evaluated on significantly longer sequences, highlighting the challenge of zero-shot generalization to unseen context lengths.}
    \label{fig:passkey_acc}
\end{figure}
\FloatBarrier

The Passkey retrieval task embeds a passkey (a four-digit number) within a long document containing unrelated text and challenges the model to retrieve it. This task serves as a simple yet intuitive benchmark for evaluating whether the model retains critical information from previous segments or disregards it in favor of new input. In this study, we used the Llama-3.2-1B model (B represents Billion) as a baseline with Condensation logic applied, and the model was continuously pretrained on the Fine-web dataset~\cite{Penedo} to adapt to the new logic with a 512-segment length. The model was then finetuned on the Passkey task dataset, starting with a sequence length of 2 K, followed by incremental increases to 4 K, 8 K, and ultimately 16 K, which was more efficient than directly finetuning to 16 K. The model was finetuned up to 16 K because of the constraints of memory resources. We evaluated the Condensation model further on 32 K and 64 K length datasets. 

Fig. \ref{fig:passkey_acc} depicts evaluation results for sequence lengths of 2 K, 4 K, 8 K, 16 K, and zero-shot evaluations at 32 K and 64 K. (Note that each results were evaluated with a 16 K length fine-tuned model) The y-axis represents accuracy, while the x-axis indicates the depth, defined as the absolute position where the passkey is embedded. The model demonstrated the ability to effectively convey information across segments, achieving 100\% accuracy within the sequence lengths on which it was fine-tuned. However, performance declined when evaluated on sequences longer than the model's fine-tuning range. Enabling the model to retrieve passkeys beyond the finetuning range in a zero-shot manner is a potential avenue for future work. 

\subsection{Summarization task}

\renewcommand{\arraystretch}{1.2}
\small
\begin{table}[h!]
\centering
\begin{tabular}{@{}lcc@{}}
\toprule
\textbf{Model (finetuned length)} & \textbf{Input Length} & \textbf{ROUGE-1/2/L (F1)} \\ \midrule
Condensation (3 K)    & 0-512                 & 0.4501 / 0.2232 / 0.3242         \\
                     & 512-1024              & 0.4368 / 0.2008 / 0.3022         \\
                     & 1024+                 & 0.3996 / 0.1510 / 0.2566         \\
                     & Overall      & \textbf{0.4292} / 0.1917 / \textbf{0.2942}\\ \midrule
Condensation (5.5 K)  & 0-512                 & 0.4532 / 0.2260 / 0.3258         \\
                     & 512-1024              & 0.4337 / 0.2002 / 0.3007         \\
                     & 1024+                 & 0.3985 / 0.1501 / 0.2557         \\
                     & Overall      & 0.4282 / \textbf{0.1918} / 0.2936 \\ \midrule
Full-Finetuned (3 K)  & 0-512                 & 0.4432 / 0.2155 / 0.3162         \\
                     & 512-1024              & 0.4271 / 0.1953 / 0.2947         \\
                     & 1024+                 & 0.3999 / 0.1650 / 0.2627         \\
                     & Overall      & 0.4231 / 0.1913 / 0.2905 \\ \midrule
Instruction model    & 0-512                 & 0.3354 / 0.1376 / 0.2171         \\
                     & 512-1024              & 0.3191 / 0.1251 / 0.1985         \\
                     & 1024+                 & 0.3022 / 0.1095 / 0.1820         \\
                     & Overall      & 0.3180 / 0.1236 / 0.1981 \\ \bottomrule
\end{tabular}
\caption{Rouge (Recall-Oriented Understudy for Gisting Evaluation) scores \cite{Lin} by model and input length. The Condensation model was further finetuned on 5.5 K to see if there was any performance improvement with a longer context. Rouge scores for each range were recorded separately to see if performance decreases with longer inputs.}
\label{tab:rouge_scores_compact}
\end{table}
\FloatBarrier
While the passkey task is designed to evaluate the model's ability to transfer information across segments, the summarization task assesses its capability to comprehend and integrate the full context. For this purpose, we scaled the model to Llama-3.2-3B and conducted continual pretraining on the Fine-web dataset~\cite{Penedo} with a 512-segment length. Subsequently, the model was first finetuned on the CNN/DailyMail dataset \cite{Karl}, which comprises a collection of news articles and their corresponding highlights. The highlights were treated as summaries, while the news articles served as the contextual input. To train the model on longer contexts, the Booksum dataset was filtered to include sequences of 5.5 K length and combined with CNN/DailyMail for the second experiment. Two baseline models were prepared for benchmarking: Llama-3.2-3B-Instruct and Llama-3.2-3B fine-tuned exclusively on the CNN/DailyMail dataset. We employed a simple generation setting with a temperature of 0.3, generating 256 tokens without beam search.

Remarkably, our model achieved higher Rouge scores (Rouge-1, Rouge-2, Rouge-L) on the CNN/DailyMail than the Llama-3.2-3B-Instruct model and comparable results to the Llama-3.2-3B model fine-tuned on the dataset in the table \ref{tab:rouge_scores_compact}. This demonstrates the model's ability to effectively condense information from each segment into limited-sized context vectors.

\FloatBarrier
\begin{figure}[h]
    \centering
    \includegraphics[width=0.45\textwidth, height=5cm]{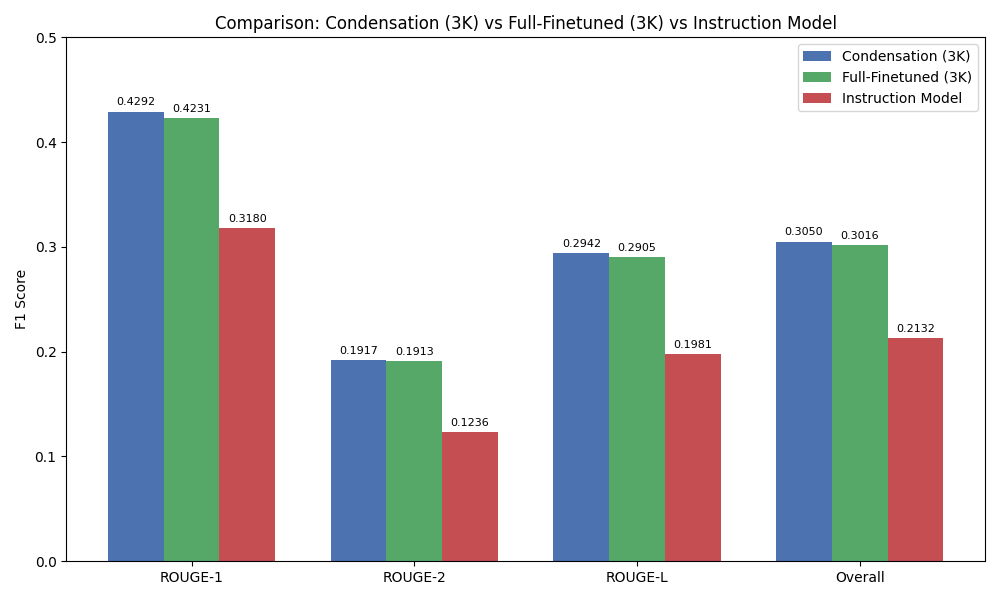}
    \caption{ROUGE score comparison between three models: (1) Condensation 3 B model fine-tuned on CNN/DailyMail with a 3K context length, (2) Llama-3.2-3B model fine-tuned on CNN/DailyMail with a 3K context length, and (3) Llama-3.2-3B Instruct model.}
    \label{fig:rouge}
\end{figure}
\FloatBarrier

In Fig. \ref{fig:rouge}, the Condensation model consistently outperforms the instruction-tuned model across all ROUGE metrics (ROUGE-1, ROUGE-2, and ROUGE-L) and achieves comparable performance to the fully fine-tuned Llama model, demonstrating its ability to effectively summarize long documents despite the segmented processing mechanism.

\begin{figure*}[t!]
    \centering 
    \includegraphics[width=1\linewidth]{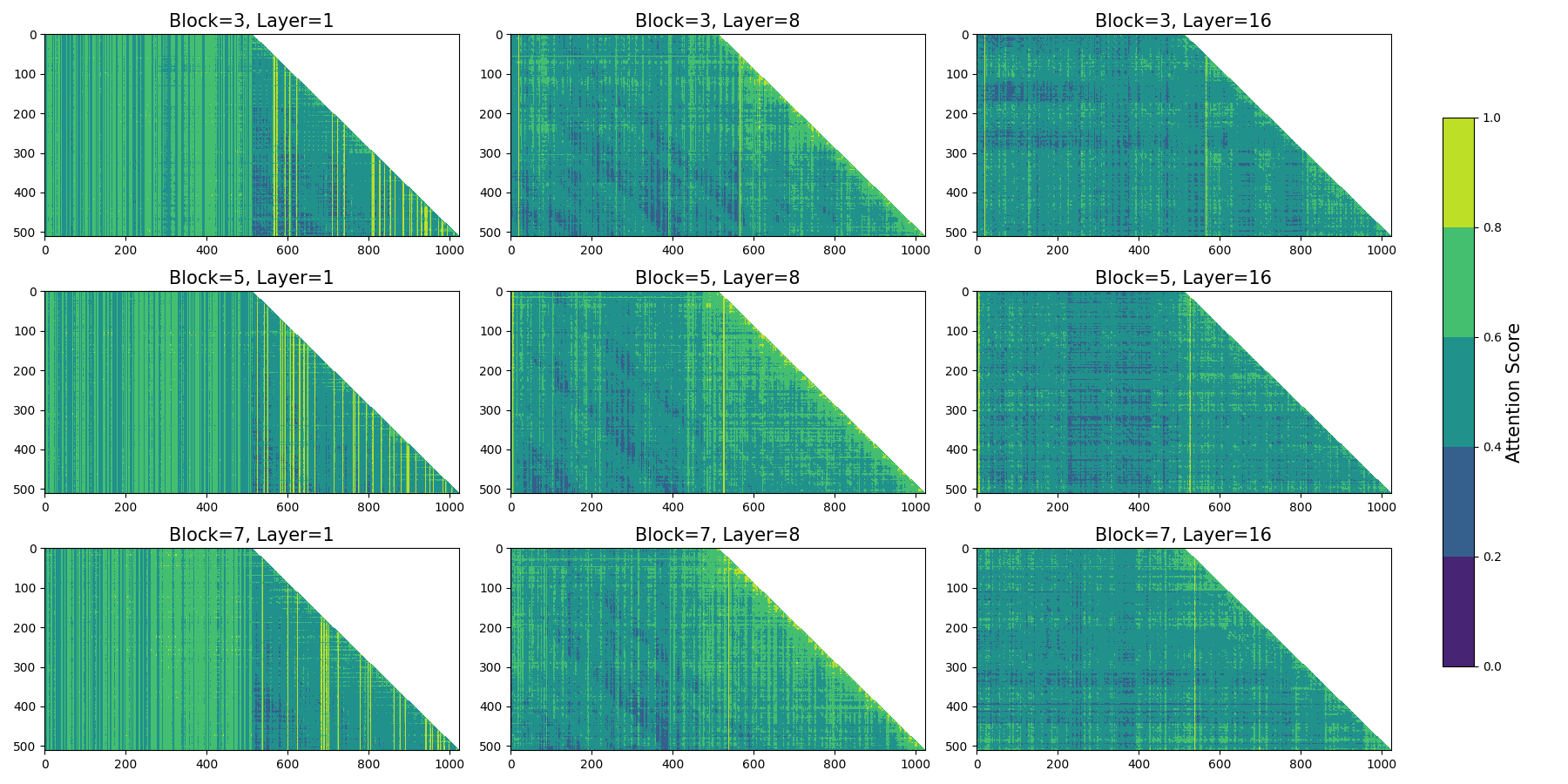} 
    \caption{Attention score of the Condensation model across different blocks and layers}
    \label{fig:contour}
\end{figure*} 

Fig. \ref{fig:contour} visualizes attention scores in a Condensation model, showing attention matrices between segments 2-3, 4-5, and 6-7 across Layers 1, 8, and 16. The input is the text in summary format with 4 K total and 0.5 K segment length, and we retrieved the attention score before applying softmax computation and rescaled with the Minmax scaling for visibility. Here we used an attention mask from Fig. \ref{fig:mask}. The matrices, with x- and y-axes ranging from 0 to 1 K and 0 to 0.5 K tokens respectively, use a color gradient (purple to yellow, 0.0 to 1.0) to represent attention weights. The visualization reveals that attention scores are well distributed across both $K^{s-1}$ and $K^{s}$, indicating that the model effectively utilizes not only the current key but also the past key when processing hidden states. The early layer (Layer 1) tends to attend previous context as much as the current segment, the middle layer (Layer 8) and later layer (Layer 16) show diffused attention capturing broader relationships, suggesting a shift from the previous to current segment context understanding as the model progresses through blocks, with slight variations in intensity across blocks.

\section{Conclusion}

In this study, we presented a novel theoretical formulation of the Transformer architecture through the framework of the Path Integral formalism, demonstrating that attention mechanisms can be reinterpreted as integrations over all potential transition paths of token states, with their temporal evolution dictated by FFNs. This reinterpretation facilitates a deeper theoretical understanding of Transformers, aligning their operation closely with the quantum mechanical principle of summation over probabilistic trajectories, each weighted by an action-dependent phase factor $\exp\left( \frac{i}{\hbar}\int^{t_{\rm f}}_{t_{0}} {\cal L} \,dt  \right) $. Here, the phase evolution and associated dynamics are encoded in a Lagrangian-like operator, and the Planck constant is appropriately scaled to match the energy scale defined by the training data, underscoring the necessity of a probabilistic interpretation for accurate future state prediction. The learned adjustment of attention weights and phase shifts during training parallels the Path Integral process of identifying classical paths. Hence, the model’s robust inference capability across diverse datasets post-training implies that the Transformer neural network inherently captures classical paths leading to optimal token prediction.

By mapping the attention and neural structures of Transformer layers to the Path Integral framework, we identified a crucial challenge: maintaining phase coherence across token states to avoid destructive interference that leads to information loss. To address this, we introduced a periodic phase gap mechanism and recurrent segment processing, ensuring that token state evolution across segments preserves historical context through constructive interference. This mechanism is mathematically enforced by the temporal alignment condition, $t_n^s = t_{n}^{s-1} + \Delta t_n$, which makes an integer-multiple phase difference between segments at each layer, ensuring continuous information propagation and long-term dependency retention. This approach can be viewed as an artificial intelligence counterpart to human short-term memory, enhancing the quality of multi-turn conversations between humans and AI systems.

Experimentally, we validated our approach using two benchmark tasks—Passkey retrieval and long-document summarization. Our method demonstrated effective contextual retention across extended sequences, while notably scaling memory usage linearly rather than quadratically with sequence length, thus significantly enhancing computational efficiency compared with traditional Transformer models. 
Empirical evaluations confirmed that our model achieved comparable or superior performance in contextual information retrieval compared with the baseline Llama-3.2 models, while significantly reducing computational resources, as quantitatively evidenced by memory consumption and training speed metrics, where the conventional Transformer required up to 2.78 times more memory and 1.68 times more training time at 4 K sequence length (4.33 times more memory at 8 K).

In conclusion, our quantum-inspired formulation provides both a deeper theoretical understanding and substantial practical improvements in Transformer architectures. By leveraging Path Integral formalism, we not only reinforce interpretability but also substantially mitigate computational bottlenecks associated with long-range dependency modeling, opening promising avenues for future research into efficient and scalable Transformer-based models.

\begin{table*}[h!]
\centering
\begin{tabular}{@{}lcccccc@{}}
\toprule
\textbf{Task} & \multicolumn{2}{c}{\textbf{Passkey}} & \multicolumn{4}{c}{\textbf{Summarization}} \\
\textbf{Configuration} & \textbf{Pretrain} & \textbf{Finetuning} & \textbf{Pretrain} & \textbf{Finetuning} & \textbf{Finetuning} & \textbf{Finetuning} \\
\midrule
Model                  & Condensation & Condensation & Condensation & Condensation & Condensation & Llama \\
Original Model         & Llama-3.2-1B & Llama-3.2-1B & Llama-3.2-3B & Llama-3.2-3B & Llama-3.2-3B & Llama-3.2-3B \\
Layer                  & 16 & 16 & 28 & 28 & 28 & 28 \\
Embedding Size         & 2048 & 2048 & 3076 & 3076 & 3076 & 3076 \\
Trainable Parameters   & 1.2 B & 1.2 B & 3.2 B & 3.2 B & 3.2B & 3.2B \\
Segment Length         & 512 & 512 & 512 & 512 & 512 & 512 \\
Total Length           & 4 K & 2 k/4 k/8 k/16 k & 5 K & 3 K & 5.5 K & 3 K \\
Learning Rate & 
\makecell{$6 \times 10^{-5}$ to \\ $2 \times 10^{-5}$} & 
$6 \times 10^{-5}$ & 
\makecell{$6 \times 10^{-5}$ to \\ $6 \times 10^{-6}$} & 
\makecell{$6 \times 10^{-6}$ to \\ $6 \times 10^{-7}$} & 
\makecell{$6 \times 10^{-6}$ to \\ $6 \times 10^{-7}$} & 
\makecell{$6 \times 10^{-6}$ to \\ $6 \times 10^{-7}$} \\
Dataset & 
\makecell{Fineweb-edu \\ (filtered over 4K)} & 
Passkey dataset & 
\makecell{Fineweb-edu \\ (filtered over 6K)} & 
CNN/DailyMail & 
\makecell{CNN/DailyMail \\ + filtered Booksum} & 
CNN/DailyMail \\
\midrule
Training Steps         & 6491 & \makecell{500/300/1000/1000}  & 9426 & 3700 & 3855 & 112 \\
Validation Loss        & 2.323 & $\simeq 0$ & 2.160 & 1.330 & 1.350 & 1.290 \\
\bottomrule
\end{tabular}
\caption{Configuration for passkey and Condensation experiments across pretraining and finetuning}
\label{table:config}
\end{table*}
\FloatBarrier
\appendices
\section{Passkey retrieval and Summarization task configuration}

Table \ref{table:config} presents the experimental configurations for pretraining and finetuning models on Passkey and Booksum tasks. For the Passkey task, a Condensation model based on Llama-3.2-1B with 16 layers and a 2048 embedding size was continuously pretrained on the Fineweb-edu dataset (filtered for sequences over 4 K) and finetuned on the Passkey dataset, achieving a validation loss of 2.323 after continuous pretraining. For the Booksum task, a larger Condensation model based on Llama-3.2-3B with 28 layers and a 3076 embedding size was continuously pretrained on Fineweb-edu (filtered over 6 K) and finetuned across two setups: (1) CNN/DailyMail, (2) CNN/DailyMail with filtered Booksum data. All experiments utilized the ZeRO-2 optimization \cite{Samyam}, cosine learning rate decay, Bfloat16 mixed precision, and a global batch size of 64. We trained all models until we observed convergence of the validation loss.

\FloatBarrier
\section{Summarization task results}

\begin{figure}[h]
    \centering
    \includegraphics[width=0.45\textwidth, height=5cm]{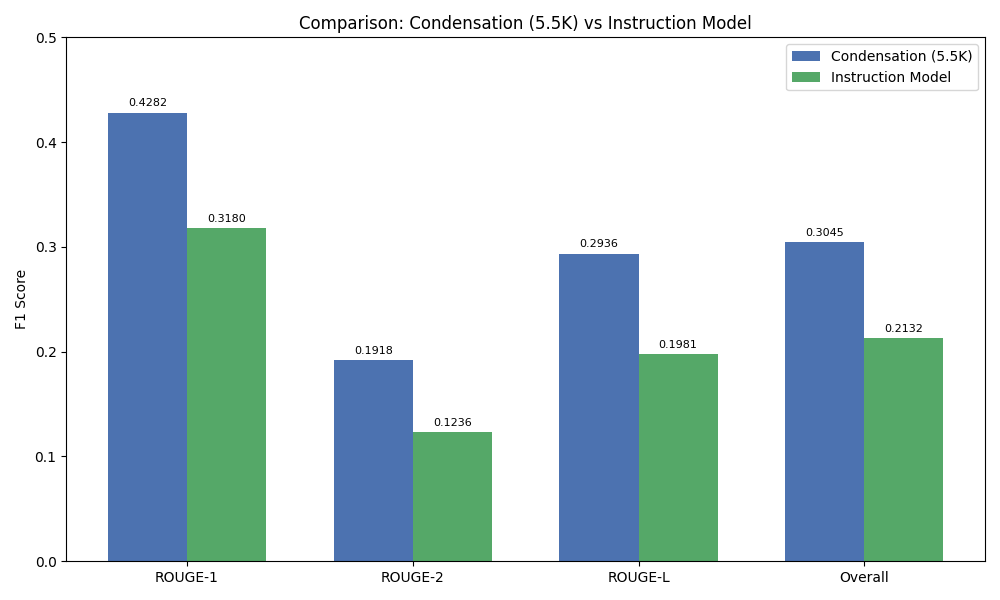}
    \caption{Comparison of ROUGE scores between two models: (1) Condensation 3 B model fine-tuned on CNN/DailyMail combined with filtered BookSum data using a 5.5 K context length, and (2) Llama-3.2-3B Instruct model. }
    \label{fig:rouge_Condensation_vs_instruction}
\end{figure}

Fig.~\ref{fig:rouge_Condensation_vs_instruction} compares the performance of the Condensation 3 B model fine-tuned on a 5.5 K context length against the Llama-3.2-3B Instruction model. Due to memory constraints, we could not train a Llama model with a 5.5 K context length for direct comparison. The results show that the Condensation model consistently outperforms the instruction-tuned baseline across all ROUGE metrics (ROUGE-1, ROUGE-2, ROUGE-L), demonstrating the effectiveness of the Condensation mechanism in handling long-document summarization tasks, even at extended context lengths.

\begin{figure}[h]
    \centering
    \includegraphics[width=0.45\textwidth, height=5cm]{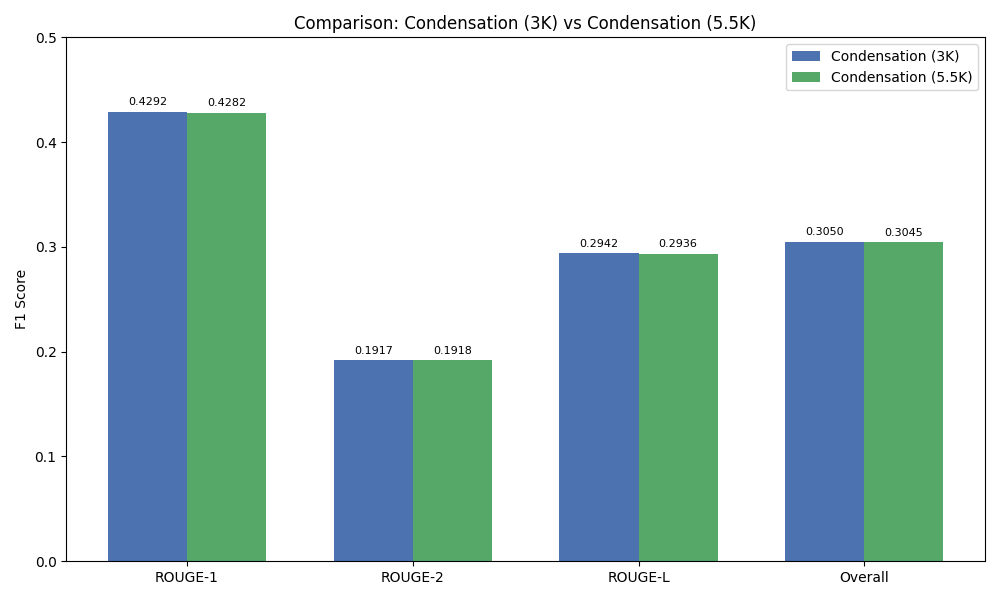}
    \caption{Comparison of ROUGE scores between two Condensation 3 B models: (1) Condensation 3B fine-tuned on CNN/DailyMail combined with filtered BookSum data using a 3 K context length, and (2) Condensation 3B fine-tuned on the same dataset with additional data using a 5.5 K context length.}
    \label{fig:rouge_Condensation_3k_vs_5k}
\end{figure}

Fig.~\ref{fig:rouge_Condensation_3k_vs_5k} presents a comparison between two Condensation 3 B models: one fine-tuned on CNN/DailyMail combined with filtered BookSum data using a 3 K context length, and the other fine-tuned on the same dataset with additional data using a 5.5 K context length. The results indicate that extending the context length to 5.5 K during training does not lead to significant performance degradation when evaluated on the CNN/DailyMail dataset.

\FloatBarrier
\section{Passkey Retrieval Task}
\label{appendix:passkey}

The following example illustrates the input format used in the Passkey retrieval task. The goal of this task is to identify and memorize a specific four-digit passkey embedded within a large amount of irrelevant text. The model must extract and recall this passkey when prompted.

\begin{center}
\begin{minipage}{0.45\textwidth}
\footnotesize
\tt
\rule{\linewidth}{0.4pt} \\
\textbf{Input:} \\

There is a critical piece of information hidden within a large amount of irrelevant text. Your task is to locate it and remember it carefully. I will test your ability to recall this information. \\

The grass is green. The sky is blue. The sun is yellow. Here we go. There and back again. (repeat $x$ times) \\

\textbf{The passkey is 9054. Remember it carefully. 9054 is the passkey.} \\

The grass is green. The sky is blue. The sun is yellow. Here we go. There and back again. (repeat $y$ times) \\

What is the passkey? The passkey is \\
\rule{\linewidth}{0.4pt}
\end{minipage}
\end{center}

\textbf{Expected Output:} 9054

\section{Summarization Task}
\label{appendix:passkey}

The following example illustrates the input format used in the Summarization task. The goal of this task is to read a lengthy piece of text and provide a concise summary of its main points when prompted.

\begin{center}
\begin{minipage}{0.45\textwidth}
\footnotesize
\tt
\rule{\linewidth}{0.4pt} \\
\textbf{Input:} \\

Summarize this text: \\

\textbf{article} \\

---- \\
Summary: \\

\rule{\linewidth}{0.4pt}
\end{minipage}
\end{center}

\textbf{Expected Output:} Summary of this article

\section*{Acknowledgment}
This work was supported in part by the High Performance Computing program through the National IT Industry Promotion Agency (NIPA) and Korea Association for ICT Promotion (KAIT).

\end{document}